%% file: roy.tex
\documentclass[graybox,natbib]{svmult}

\usepackage{mathptmx}       
\usepackage{helvet}         
\usepackage{courier}        
\usepackage{type1cm}        
\usepackage{makeidx}         
\usepackage{graphicx}        
\usepackage[bottom]{footmisc}
\usepackage[T1]{fontenc}

\usepackage{aas_macros}

\makeindex             

\def\atlas3d{ATLAS$^{\rm 3D}$}
\def\lsim{\mathrel{\rlap{\lower3.5pt\hbox{\hskip0.5pt$\sim$}}
    \raise0.5pt\hbox{$<$}}}                
\def\gsim{~\rlap{$>$}{\lower 1.0ex\hbox{$\sim$}}}

\begin{document}

\title*{Early type galaxies and structural parameters from ESO public survey KiDS}
\author{N.~Roy, N.~R.~Napolitano, F.~La~Barbera, C.~Tortora,
F.~Getman, M.~Radovich, M.~Capaccioli,\and the KiDS collaboration}
\institute{N.~Roy \email{roy@fisica.unina.it}, M.~Capaccioli \at
Dipartimento di Scienze Fisiche, Università di Napoli Federico II,
Compl. Univ. Monte S. Angelo, 80126 - Napoli, Italy \and
C.~Tortora, F.~La~Barbera, N.R.~Napolitano, F.~Getman \at INAF --
Osservatorio Astronomico di Capodimonte, Salita Moiariello, 16,
80131 - Napoli, Italy \and M.~Radovich \at INAF -- Osservatorio
Astronomico di Padova-vicolo Osservatorio 5 - 35122 Padova,Italy
\and KiDS collaboration:http://kids.strw.leidenuniv.nl/team.php}

\authorrunning{Roy N. et al.}

\maketitle

\abstract{The Kilo Degree survey (KiDS) is a large-scale optical
imaging survey carried out with the VLT Survey Telescope (VST),
which is the ideal tool for galaxy evolution studies. We expect to
observe millions of galaxies for which we extract the structural
parameters in four wavebands (u, g, r and i). This sample will
represent the largest dataset with measured structural parameters
up to a redshift $z=0.5$. In this paper we will introduce the
sample, and describe the 2D fitting procedure using the 2DPHOT
environment and the validation of the parameters with an external
catalog.}

\subsection*{Introduction}
The Kilo Degree Survey (KiDS) is a ESO public survey (PI. K.
Kuijken) carried out with the VLT Survey Telescope (VST), at the
ESO Paranal Observatory. It will cover 1500 square degrees of the
night sky in four optical bands (\emph{ugri}). The large area
covered, the depth, the good seeing and pixel scale make the VST
the ideal tool to investigate the evolution of galaxies across the
last billions of years. KiDS possess upper hand in different
aspects compared to the previous surveys. E.g., with respect to
the Sloan Digital Sky Survey (SDSS), which is the most successful
survey in the galaxy evolution studies \cite{SDSS-galaxyevol},
KiDS has two major improvements that are crucial for its science
goals: it is much deeper (by about 2 magnitudes), and it has
better image quality, particularly in the r-band with respect to
SDSS.

We will study the structural properties of early-type galaxies,
which, being the oldest and most massive systems in the universe,
provide a unique way to trace the galaxy evolution and formation,
and important physical processes like galaxy mergings.

\subsection*{The sample} For the current study we are using a galaxy
catalog extracted from the first 158 square degrees of the KiDS
Survey. Single band sourcelist have been extracted with a
stand-alone procedure for the catalog extraction optimized for
KiDS whose backbone is constituted by S-Extractor
\cite{sextractor_96} software for the source detection and
extraction which is fed with KiDS image, weighting maps (from
Astro-Wise pipeline, \cite{Astrowise_11}) and external masks for
bad pixels, saturated stars and stars haloes obtained with an
automated procedure (i.e., an evolution of the ExAM software, see
\cite{Pulcenella}). Star/galaxy separation is based on the {\tt
CLASS\_STAR} (star classification) and S/N (signal-to- noise
ratio) parameters provided by S-Extractor (\cite{sextractor_96};
\cite{2DPHOT}). From the whole galaxy population extracted within
the KIDS DR2 ($\sim7$ million), we have selected those galaxies
with r-band signal-to-noise ratio (S/N)>50. This criterion allows
one to obtain reliable structural parameters \cite{SPIDER-I}.
Using this cut on S/N ratio we ended up with a sample of
$\sim38000$ galaxies. For further details, see the contribution
from Tortora et al.

\subsection*{Structural parameter extraction} The structural
parameter extraction is performed by 2DPHOT \cite{2DPHOT}, which
is an automated software environment which accomplishes several
tasks such as catalog extraction (using S-extractor,
\cite{sextractor_96}),  star/galaxy separation and surface
photometry, for the selected galaxies. Galaxy images were fitted
with a PSF convolved S\'ersic models \cite{Sersic} having
elliptical isophotes plus a local background value.
\begin{figure}
\centering
\includegraphics[scale=.4]{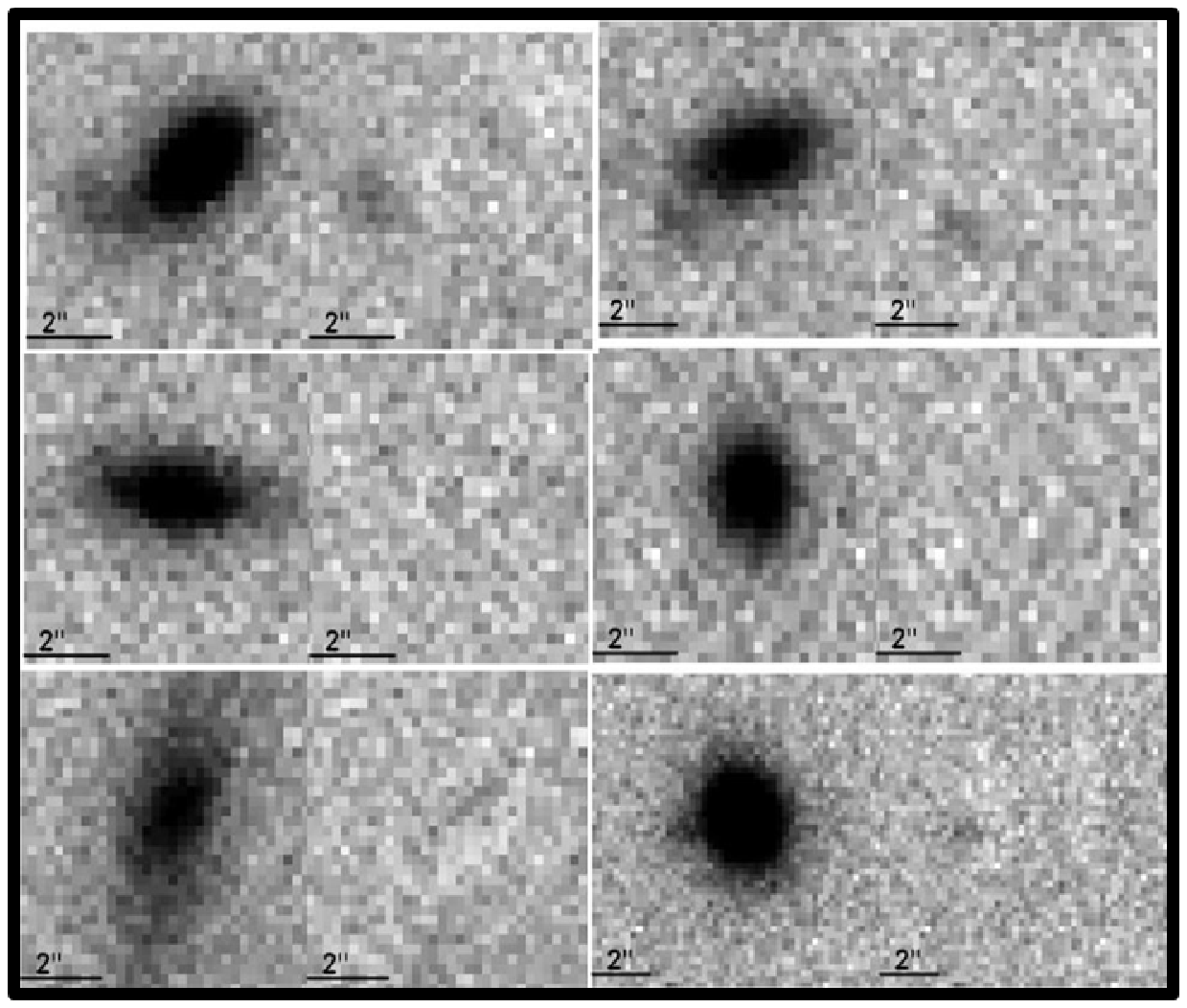}
\includegraphics[scale=.4]{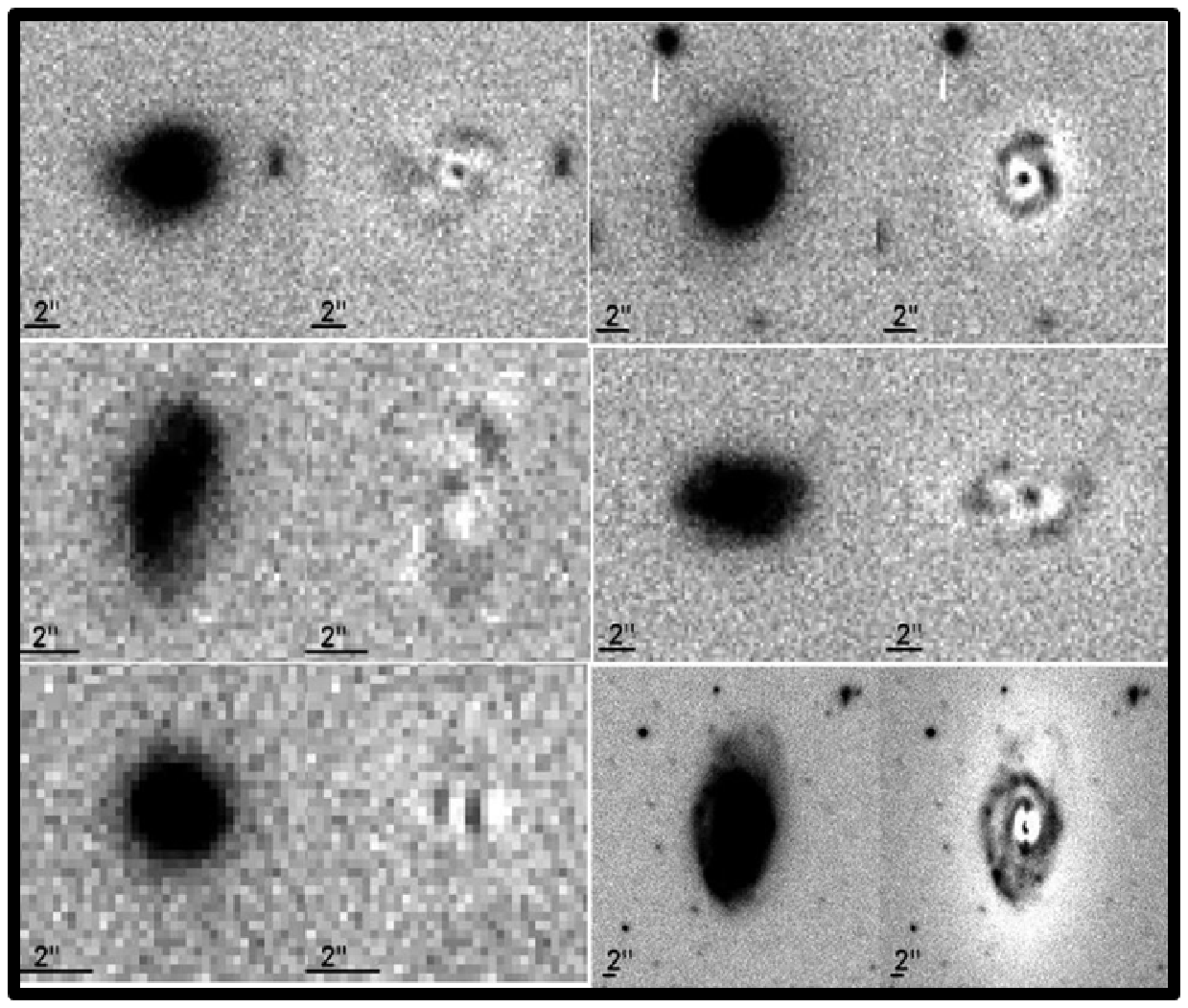}
\caption{2D fit results for 12 example galaxies in r-band. Each
frame shows the galaxy stamp (left) and the residual map (right)
after model subtraction. Left panels  are for $\chi^{2}$ < 1.5,
while right panels for $\chi^{2}$ > 1.5.} \label{fig:1}
\end{figure}
In Fig.~\ref{fig:1} we show the 2DPHOT output stamps for few
example galaxies. 93\% of our galaxy sample were fitted by a
S\'ersic profile with a small reduced $\chi^{2}$ value (see left
panels in Fig. \ref{fig:1}) without any residuals. However, we
found galaxies with $\chi^{2}$ > 1.5, not perfectly fitted, and
leaving in the residual maps signs of substructures, as spiral
arms. 2DPHOT model fitting will provide structural parameters like
surface brightness $\mu_{e}$, effective radius $R_{e}$, S\'ersic
index $n$, model magnitude, axial ratio, position angle, etc.

Because our sample consists of both early-type (ETGs) and
late-type (LTGs) galaxies, we have selected ETGs on the base of
the following criteria: 1) S\'ersic index n > 2.5, since LTGs are
found to have lower S\'ersic indices (e.g. \cite{Tortora_sersic}),
2) $R_{e}$ > 0.2 arcsec (since the VST pixel scale is 0.21
arcsec/pixel), and 3) $\chi^{2}$ < 1.5, to take the galaxies
best-fitted by the S\'ersic profile and remove those systems with
a clear sign of spiral arms in the fit residuals (see right panels
in Fig. \ref{fig:1}).
\subsection*{Comparison with literature data} To study the
reliability and consistency of the derived structural parameters
we made a comparison of our results with an external datasample
from the SPIDER survey \cite{SPIDER-I}. This dataset, extracted
from SDSS, consists of  $\sim 39000$ ETGs with redshifts 0.05 < z
< 0.095, with structural parameters determined by fitting a
S\'ersic profile and the fitting procedure is performed using
2DPHOT. We found 345 shared sources by cross-matching the two
catalogs.
\begin{figure}[h]
\centering
\includegraphics[scale=.37]{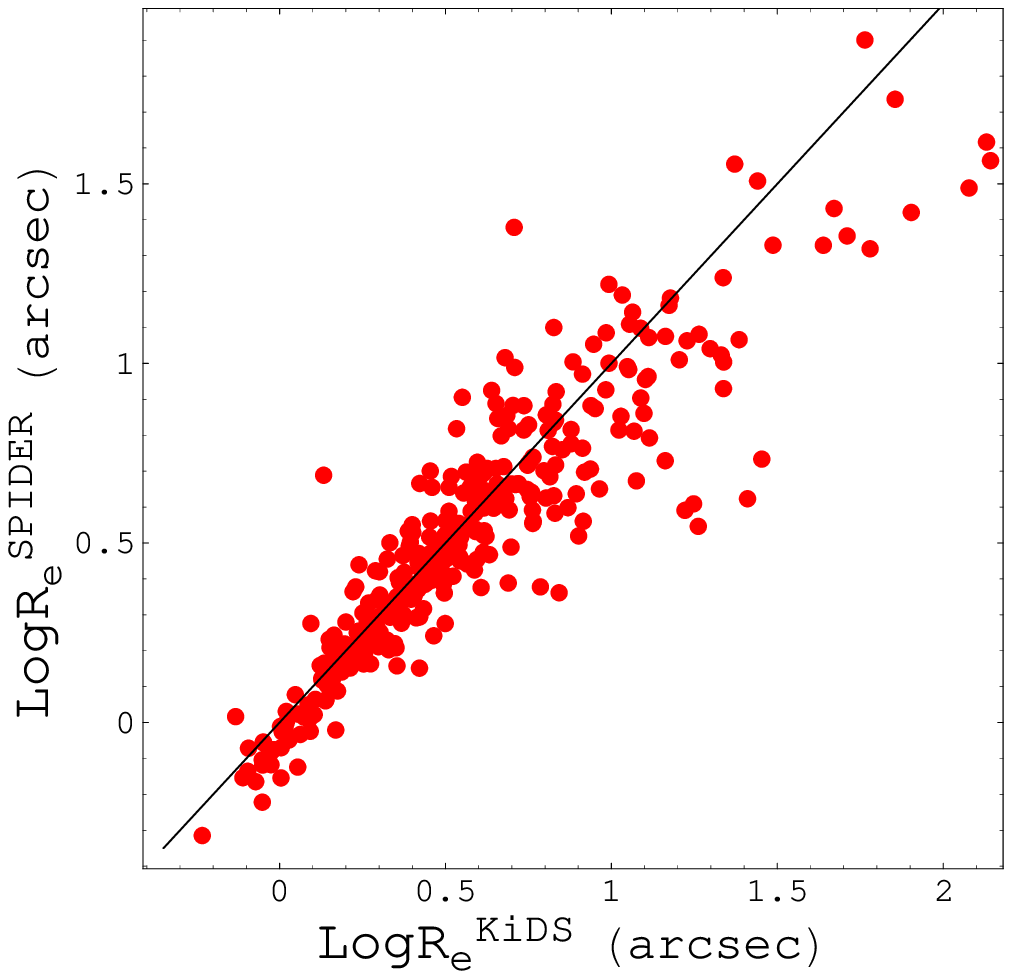}
\includegraphics[scale=.37]{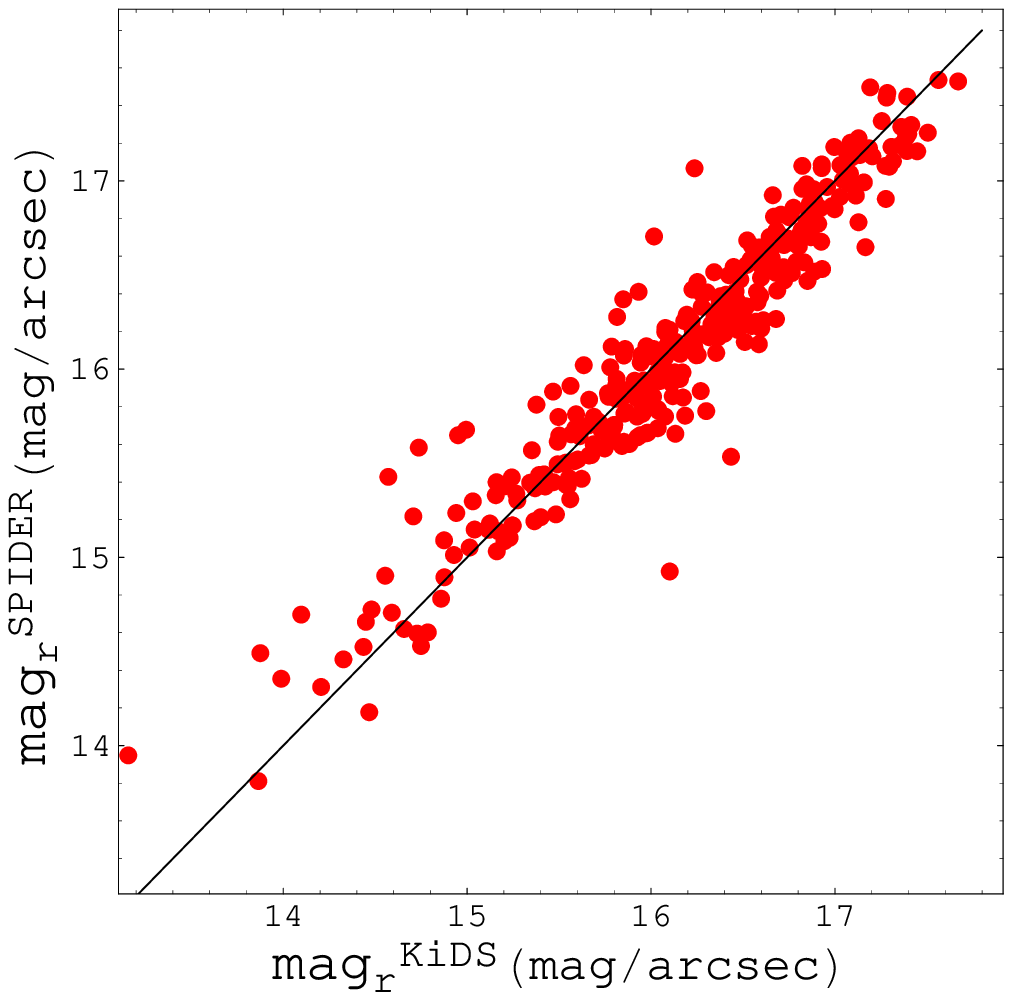}
\includegraphics[scale=.37]{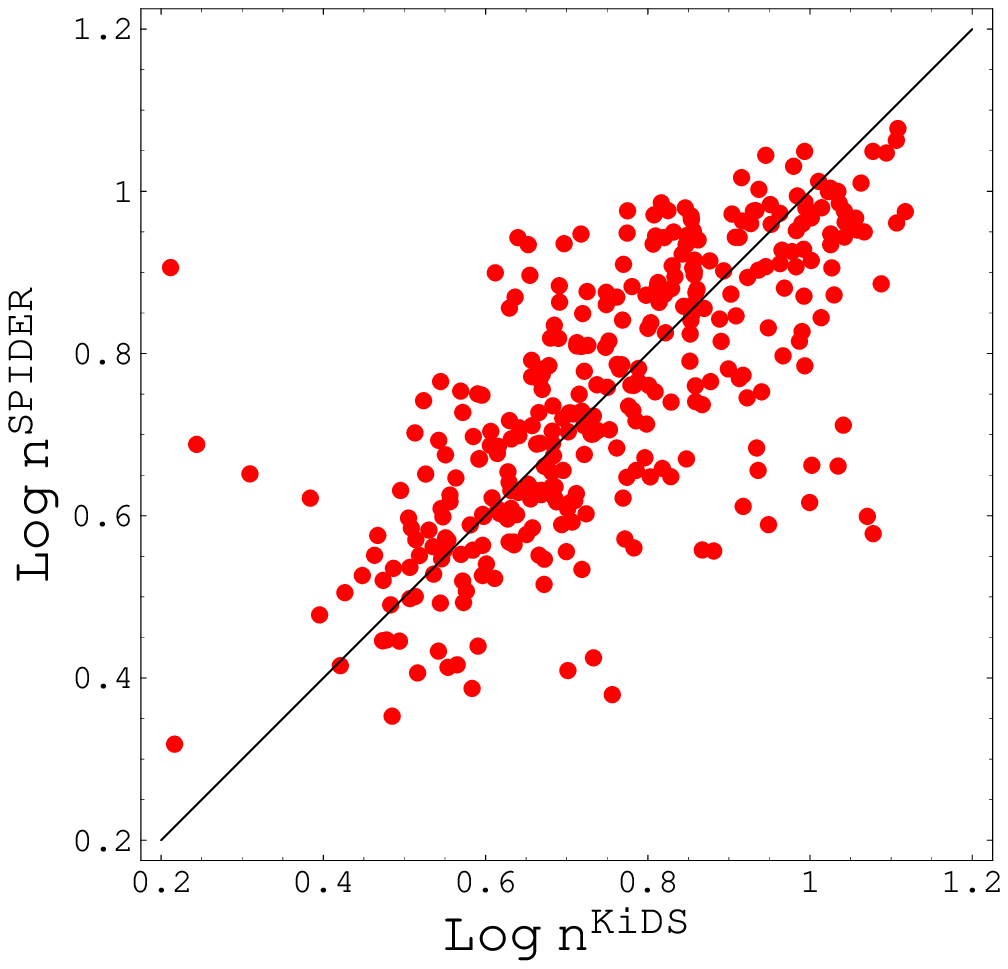}
\caption{Comparison of KiDS structural parameters with the ones
derived within the SPIDER survey. The SPIDER dataset consists of
ETGs with redshifts in the range 0.05 < z < 0.095, selected from
SDSS; the structural parameters are derived using 2DPHOT. We show
the relationship for effective radius ($R_{e}$), model magnitude
($mag_{r}$) and Sérsic index (n) respectively from left to right .
Data are shown as red points. The diagonal lines are the
one-to-one relations, corresponding to the perfect agreement.}
\label{fig:2}
\end{figure}

The comparisons of $R_{e}$, magnitudes and S\'ersic index are
shown in Fig.~\ref{fig:2}. There is a good agreement in the
parameters from two catalogs, which confirm the reliability of our
extracted parameters in the KiDS survey.

\subsection*{Conclusions}
KiDS, being a large scale survey with good image quality, provides
a great platform for the study of galaxy evolution. With the
extraction of structural parameters we are about to start the
analysis of scaling relations, as the relationship between the
extracted structural parameters (e.g. $R_{e}$ and n) as a function
of luminosity or stellar mass \cite{Trujillo_size_evol}, or the
correlation of surface brightness with $R_{e}$ or magnitude
\cite{LaBarbera_Kormendy}. KiDS will allow to analyze these
correlations at different cosmic epochs up to redshift 0.5. Galaxy
size and mass evolution with cosmic time provide fundamental
details about the physical mechanisms which drive the galaxy
evolution, including the effect of mergings or other phenomena
related to the galaxy environment.

\input{roy_referenc}

\end{document}

%% file: roy_referenc.tex
\bibliographystyle{plain}